\documentclass[10pt]{iopart}

\begin{document}
\voffset=2cm

\title[$T$-violation tests for relativity principles]{$T$-violation tests for relativity principles}

\author{Yvonne Y. Y. Wong
\footnote[3]{To whom correspondence should be addressed
(ywong@physics.udel.edu)} }

\address{Department of Physics and Astronomy, University of Delaware, Newark, DE 19716, USA}

\begin{abstract}
We consider the implications of a violation of the equivalence
principle or of Lorentz invariance in the neutrino sector for the
$T$-asymmetry $\Delta P_T \equiv P(\nu_{\alpha} \to \nu_{\beta}) -
P(\nu_{\beta} \to \nu_{\alpha})$ in a three-flavour framework.  We
find that additional mixing due to these mechanisms, while obeying
all present bounds, can lead to a substantial enhancement,
suppression, and/or sign change in $\Delta P_T$ for the preferred
energies and baselines of a neutrino factory. This in turn allows
for the possibility of improving existing constraints by several
orders of magnitude.
\end{abstract}




\section{Introduction}
The phenomenon of flavour oscillations follows from non-degenerate
neutrino states and mixing.  The ``standard'' mechanism stipulates
that the degeneracy be broken by means of small neutrino masses.
An alternative scenario that does not require neutrino masses is
provided by a violation of the equivalence principle (VEP) in the
neutrino sector \cite{bib:gasperini,bib:halprin}.

Phenomenologically, the standard and the VEP mechanisms differ in
their energy dependences.  In the two-flavour case, a formal
transformation from the former to the latter scenario is
accomplished by replacing in the oscillation probability $\delta
m^2/2 E \to 2 E |\phi| \delta \gamma$, where $\delta \gamma \equiv
\gamma_2 - \gamma_1$ is the difference in the neutrino--gravity
couplings, and $\phi$ is the gravitational potential.  Note that a
violation of Lorentz invariance also contributes to lifting the
neutrino degeneracy \cite{bib:liv}. This scenario, however, has
the same phenomenology as VEP \cite{bib:vepliv}.

The purpose of this work is not to promote  VEP  as a solution to
the solar, atmospheric, and LSND neutrino puzzles---this has been
shown by many to be futile (see \cite{bib:cpvep} for a review),
and the VEP breaking scale
 is severely constrained by null oscillation
experiments, $|\phi|\delta \gamma \!< \! 10^{-22}$
\cite{bib:pantaleone}. Rather, we examine the possibility of
further constraining $|\phi| \delta \gamma$ with a new generation
of experiments, namely, the measurement of the $T$-asymmetry
$\Delta P_T  \equiv  P(\nu_{\alpha}\to\nu_{\beta}) - P(\nu_{\beta}
\to \nu_{\alpha})$ at a neutrino factory.

\section{$T$-violation}

Consider three-flavour oscillations governed by the Hamiltonian
(in flavour basis)
\begin{equation}
\label{eq:totalham} \widetilde{H}  = \frac{1}{2E} U_M M
U_M^{\dagger} + 2 E \ U_G G U_G^{\dagger} + V \equiv H^M + H^G +
V,
\end{equation}
where $M \!=\! {\rm Diag} (- \delta m^2_{\odot},  0,  \delta
m^2_{\oplus} )$ is the mass term, $G \!=\! |\phi|\! \times\! {\rm
Diag}(\gamma_1,  \gamma_2,  \gamma_3)$ comes from VEP, $V\! =\!
\sqrt{2}\  G_F N_e \!\times\! {\rm Diag}(1,  0,  0)$ represents
matter effects, $U_M$ is the MNS matrix with three angles and a
$CP$ phase $\delta$, and $U_G$ describes mixing in the VEP sector.
We assume the solar and atmospheric neutrino anomalies to be
completely explained by the square mass differences $\delta
m^2_{\odot}$ and $\delta m^2_{\oplus}$, together with the angles
$\theta_{\odot}$ and $\theta_{\oplus}$ in $U_M$.

 The asymmetry between the $T$-conjugate
oscillation probabilities is given by
\begin{equation}
\label{eq:deltapt} \Delta \widetilde{P}_{T}  = 16 \widetilde{J} \
\sin \frac{\widetilde{\Delta}_{12} L}{2}\ \sin
\frac{\widetilde{\Delta}_{23} L}{2}\ \sin
\frac{\widetilde{\Delta}_{31} L}{2},
\end{equation}
with $\widetilde{\Delta}_{ij} \!= \!\widetilde{\lambda}_i \! -\!
\widetilde{\lambda}_j$, where  $\widetilde{\lambda}_{1,2,3}$ are
the eigenvalues of $\widetilde{H}$, and $\widetilde{J}$ is the
effective Jarlskog factor.  We wish to compare $\Delta
\widetilde{P}_T$ with the intrinsic  $T$-asymmetry $\Delta P_T$
due to $\delta$. Using the identity
 $\Delta_{12} \Delta_{23}
\Delta_{31} J\! = \! {\rm Im}(H_{e \mu} H_{\mu \tau } H_{\tau e})$
\cite{bib:naumov,bib:harrison}, and by restricting the neutrino
energy and the experimental baseline to ${\cal O}(10) \!\!\to\!\!
{\cal O}(30) \ {\rm GeV}$  and ${\cal O}(3000)\! \! \to\!\!  {\cal
O}(6000)\ {\rm km}$ respectively, we find the ratio $\Delta
\widetilde{P}_T/\Delta P_T$ to be well approximated by
\begin{equation}
\label{eq:ratioapprox} \frac{\Delta \widetilde{P}_T}{\Delta P_T}
\simeq  \frac{{\rm Im}(\widetilde{H}_{e \mu} \widetilde{H}_{\mu
\tau } \widetilde{H}_{\tau e})}{{\rm Im}(H_{e \mu} H_{\mu \tau }
H_{\tau e})},
\end{equation}
where $H \equiv H^M + V$ is the Hamiltonian (\ref{eq:totalham})
without VEP contribution, assuming that $|\phi| \delta \gamma$
satisfies the condition $2 E |\phi \delta \gamma | < |\delta
m^2_{\oplus}|/2E$.

\section{An example}

The exact consequence of VEP on the $T$-asymmetry depends on the
nature of $H^G$.  We provide here the simplest example of pure
$\nu_e \leftrightarrow \nu_{\tau}$ mixing in the VEP sector, with
VEP breaking scale $|\phi| \delta \gamma$ and mixing angle
$\varphi$. Equation (\ref{eq:ratioapprox}) can now be written as
\begin{equation}
\frac{\Delta \widetilde{P}_T}{\Delta P_T} \simeq 1 - \frac{4
E^2}{\delta m^2_{\odot}} \frac{\sin 2 \theta_{\oplus}}{\sin 2
\theta_{\odot}} |\phi| \delta \gamma \sin 2 \varphi.
\end{equation}
Using the best fit values $\delta m^2_{\odot} \simeq 5 \times
10^{-5}\ {\rm eV}^2$, $\sin 2 \theta_{\odot} \simeq 0.8$, and
$\sin 2 \theta_{\oplus} \simeq 1$, we see that at $E = 13\ {\rm
GeV}$, the VEP parameters  $|\phi| \delta \gamma \sin 2 \varphi=
\pm 10^{-25}$ can offset the $T$-asymmetry by $\mp 100\ \%$, i.e.,
$\Delta P_T$ is enhanced or suppressed depending on the sign of
$|\phi| \delta \gamma \sin 2 \varphi$. Furthermore, if the offset
exceeds $100\ \%$---because of a large VEP breaking scale and/or a
high neutrino energy, $\Delta P_T$ flips sign.

\section{Conclusion}

Extra neutrino mixing from VEP can enhance, suppress, and/or flip
the sign of the intrinsic asymmetry between two $T$-conjugate
oscillation processes, depending on the VEP breaking scale and the
neutrino energy. Thus $T$-violation experiments run at two or more
energies will allow us to either establish, or further constrain
VEP and improve present bounds on $|\phi| \delta \gamma$ by at
least two orders of magnitude.

\end{document}